\documentclass[showpacs,showkeys,twocolumn]{revtex4}

\usepackage{amsmath}
\usepackage{amssymb}
\usepackage{graphicx}
\usepackage{natbib}
\usepackage{hyperref}

\begin{document}

\title{A Database for Fast Access to Particle-Gated Event Data}

\date{August 12, 2006}

\author{A.~B\"urger}
\email{buerger@hiskp.uni-bonn.de}
\affiliation{Helmholtz-Institut f\"ur Strahlen- und Kernphysik,
  Nu\ss{}allee 14-16, D-53115 Bonn, Germany}

\begin{abstract}
  In nuclear physics experiments involving in-flight fragmentation
  of ions, usually a large number of different nuclei is produced
  and various detection systems are employed to identify the species
  event by event, e.g. by measuring their specific energy loss and
  time-of-flight.
  For such cases -- not necessarily limited to nuclear physics --
  where subsets of a large dataset can be identified using a small
  number of measured signals a software for fast access to varying
  subsets of such a dataset has been developed.
  The software has been used successfully in the analysis of a one
  neutron knock-out experiment at GANIL.
\end{abstract}
  
\pacs{01.50.hv, 07.05.Kf}
\keywords{data analysis}

\maketitle

\section{Introduction}
\label{sec:intro}

During nuclear physics experiments, data are usually stored in the
so-called ``list-mode'', that is in the sequence of their occurrence
which will be called ``time-ordered'' in this article.
This is the most natural if not the only reasonable way to store
measured data during the experiment.
But the analysis of time-ordered data might not be efficient if the
interesting nuclei represent only a small portion of the events in the
dataset.
In this case, the analysis of time-ordered data means that, to select
a nucleus of interest with varying conditions, the whole dataset has
to be read with the majority of the data being ignored after finding
out that they do not match the selection criteria.
As all data have to be read, the analysis time is proportional to the
size of the whole dataset, independent of how much of the data are
actually used.
The aim of the work presented here was to develop a software that
allows to store event data in a way that allows to reduce the amount
of data read unnecessarily and thereby to reduce analysis time.

Although the software has been developed with particle-identification
in nuclear physics experiments in mind, it is not restricted to this
type of data.

Note that for high-multiplicity $\gamma$-ray data, Cromaz
et~al.~\cite{Cro01} have developed {\em Blue}, a database program that
sorts the events by $\gamma$-ray energies.
It allows fast access to events identified by a set of $\gamma$-ray
energies.
The {\em Blue} database is based on a very similar principle but
focused and specialised on high-multiplicity $\gamma$-ray data.

The standard way to improve the speed of analysis is to do a
``pre-sorting'':
the time-ordered dataset is split ``manually'' into different subsets,
e.g. into one subset for each nucleus.
Using the sorting software presented here has some advantages over
this technique:
the work of dividing into subsets is done mostly automatic -- only
quantities suitable for selection have to be calculated by user code
-- and all data are stored in one file for convenient handling and
interactive access.
Furthermore also the selection criteria are flexible, as changing them
does not require re-reading the time-ordered dataset.

\section{Sorting and indexing algorithm}
\label{sec:descr}

To reduce the number of data being read unnecessarily, the data are
partially ordered and indexed with respect to some quantities $q_k$.
The $q_k$ might, e.g., represent time-of-flight or energy-loss values
allowing to identify particles.
These values $q_k \in \mathbb{Q}_k$ with $1\le k \le m, m\ge2$ have to
be extracted from the data by user code.
The $\mathbb{Q}_k$ denote the sets of values for the $q_k$, e.g. sets
of integer and floating-point numbers.
The sorting procedure divides the space $\mathbb{Q}_1 \times \cdots
\times \mathbb{Q}_m$ into sub-volumes -- they are called ``buckets''
here -- which contain events with similar values of the $q_k$.
The dimensions of each ``bucket'' sub-volume are stored in an index
when the sorting process is finished.
When reading the dataset applying selection criteria on the $q_k$, the
index is used to identify the buckets which have to be looked at.
Only the events in those buckets are actually read -- no time is spent
reading the events in the other buckets, as they are then known not to
match the selection criteria.

\begin{figure}
  \centering
  \includegraphics[width=6cm]{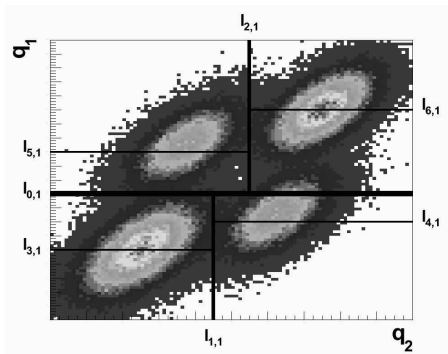}
  \caption{%%
    Schematic illustration of the construction of the bucket tree
    for $m=2$ and $n_l=1$ with $q_1$ on the ordinate and $q_2$ on
    the abscissa.
    Each line represents the limit $l_{b,1}$ for the two child
    buckets.
    Thicker lines indicate a lower depth.
  }
  \label{fig:tree}
\end{figure}

In more detail, the sorting procedure works as follows.
A user code reads events from the list-mode storage and calculates the
$m$ quantities $q_k$ event by event.
Each event is then handed over to the sorting program which sorts them
into a tree structure of buckets.
The bucket tree is built while the time-ordered dataset is read.
In the beginning there is just one bucket with number $b=0$ at depth
$d=0$ which represents the full space $V_{b=0}=\mathbb{Q}_1 \times
\cdots \times \mathbb{Q}_m$.
The program starts putting events into it until it contains $n_s$
events, the user-defined maximum number of events a bucket may
contain.
Whenever a bucket number $b$ with depth $d$ contains $n_s$ events, the
bucket is split and assigned $n_l+1$ child buckets with numbers $b_i$
and depth $d+1$ whose volumes are
\begin{displaymath}
  V_{b_i} = V_b \cap \tilde V_{b_i}
\end{displaymath}
with
\begin{multline*}
  \tilde V_{b_i} = \mathbb{Q}_1 \times \cdots \times
  ([l_{b,i-1},l_{b,i}] \cap
  \mathbb{Q}_k) \cdots \times \mathbb{Q}_m, \\
  i=1,2,\dots,n_l+1, \; l_{b,0} = -\infty, \; l_{b,n_l+1} = +\infty.
\end{multline*}
Here, $k=(d \bmod m)+1$ so that with increasing depth e.g. for $m=2$
the volumes are divided alternatingly along the $q_1$ and $q_2$ axes
(see fig.~\ref{fig:tree}).
The limiting values $l_{b,i}$ for the child buckets of $b$ are
calculated from an analysis of the $q_k$ values of the events in
bucket $b$.
After creating the child buckets, as the last step of splitting, the
events from the full bucket $b$ are sorted into the child buckets
$b_i$.
Applying this procedure repeatedly, a tree of buckets is created where
each bucket at depth $d+1$ contains events or has children with a more
limited range in the value $q_k$ (with $k=(d \bmod m)+1$) than the
events in the parent bucket at depth $d$.

During the construction of the tree, the data of all buckets have to
be kept in memory for fast splitting.
As the amount of available memory will not be sufficient to keep all
events, the procedure is stopped when a predefined number of buckets
$n_b$ is created: the splitting of buckets is stopped and events are
immediately sorted into the appropriate buckets and written to disk.
The numbers $n_b$, $n_s$ and $n_l$ have to match the amount of
available memory.
In the example described below, $n_b=1024$, $n_s=16$ and $n_l=1$ were
used and for simplicity the single limit was determined as
\begin{displaymath}
  l_{b,1}=\tfrac{1}{2}(\min(q_k)+\max(q_k))
\end{displaymath}
for the $q_k$ of the events in the bucket being split.
Finally, when the end of the input is reached, an index is written
where all the limits $l_{b,i}$ are registered.
The file obtained this way contains the events partially sorted by the
$q_k$.
The depth and the size of the buckets may vary widely and will reflect
the distribution of the sorted quantities $q_k$, especially if these
quantities are correlated.

As the tree structure is built up from a subset of the events, it is
important that the distribution of the events is the same throughout
the dataset.
Feeding data already sorted by any of the $q_k$ into the program will
produce a database without much or even without any gain in access
speed.
Neither does it make sense to select a monotonously increasing value
as one of the ordering quantities $q_k$.
Furthermore, the sorting requires that the data are in the form of
self-contained units (events) -- as they are re-ordered, any
information contained in their order is lost.

If the granularity of the buckets is not sufficient after the first
sorting run, the number of buckets can be increased using a second
program that builds a new tree for each of the childless buckets in
the input database.
It works very similar to the initial sorting program, except that it
reads from a different source -- database buckets instead of the
time-ordered dataset -- and creates child buckets of the input bucket.
Again the required constancy of the distribution of the events is
used.

For a query with constraints on the desired range of one or more
$q_k$, only those buckets need to be looked at where the limits --
which are read from the index -- overlap with the desired $q_k$
ranges.
Therefore, -- depending on the query -- only a part of the buckets and
thus a portion of the data whose size is approximately proportional to
the number of selected events has to be read.
This can be much faster than reading the whole list-mode dataset,
especially if the selected $q_k$ region contains only a small number
of events.

\section{Implementation}
\label{sec:impl}

The present implementation of the sorting code, written in the
\texttt{C++} programming language, is flexible regarding the structure
of the events that are to be stored:
data of integer (signed and unsigned 1-, 2- or 4-byte) or floating
point type can be stored in almost any order and with almost arbitrary
names for later interactive retrieval.
Arrays of values can be defined with either fixed or variable size,
e.g. to accommodate $\gamma$-ray data with varying multiplicity.
To reduce the size of the produced database, the data are compressed
using the standard compression library {\tt zlib}~\cite{zlib}.
An easy to use interface to ROOT~\cite{Bru97} with a query format
similar to {\tt TTree::Draw} can be used to interactively produce
histograms and matrices.
This interface allows using the ROOT graphical cuts to select pairs of
the $q_k$.
Note that ROOT trees can store data having a tree-like event structure
(which may be much more complex than the currently implemented event
structures of the database sorting program), but that ROOT trees,
being a general and very efficient list-mode storage format, do not
have an index for data access as described here.
For non-interactive queries, it is possible to use a {\tt C++}
interface to access the stored data.

The database can be extended with ``friends'', these are files
containing additional data for all the events in a database.
For example, $\gamma$-ray data produced with different calibration
coefficients might be stored in different friend files, so that the
calibrations can be reviewed without making copies of the data for the
other detectors.
It is also possible to have friends calculating values on-the-fly,
e.g. lookup of detector positions if only the detector number has been
stored.
The use of these friends can reduce the size of the database
significantly.

\section{Example}
\label{sec:perf}

\begin{figure}[tb]
  \centering
  \includegraphics[width=6cm]{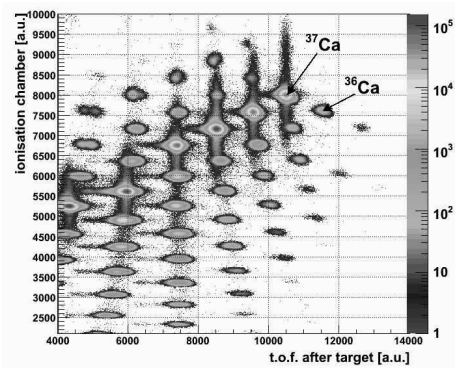}
  \caption{%
    Typical particle identification matrix obtained in the GANIL
    experiment~\cite{Bur06}.
    On the abscissa, the value of the time-of-flight measurement for
    $A/Q$ identification is shown, on the ordinate the energy loss
    measured in the ionisation chamber for the charge identification,
    both in arbitrary units.
    Each blob corresponds to one nuclear species.  }
  \label{fig:pid}
\end{figure}
The software has been developed and is being used for the analysis of
data taken in an experiment at GANIL.
The aim of the experiment was to determine the energy of the first
$2^+$ state in $^{36}$Ca by observation of the $\gamma$-rays from the
$2^+ \rightarrow 0^+$ transition~\cite{Bur06}.
In the experiment a $^{40}$Ca primary beam of $95$\,MeV/u was
fragmented in a primary carbon target in the SISSI device, producing a
cocktail of nuclear species.
Out of these products, $^{37}$Ca along with -- due to insufficient
separation -- some other nuclear species, was selected by the $\alpha$
spectrometer of GANIL, resulting in a secondary ``$^{37}$Ca'' beam.
On a secondary Be target, the $^{37}$Ca beam underwent
few-nucleon-removal, again producing many different nuclei with
$^{36}$Ca among them.
These nuclei were identified using the spectrometer SPEG~\cite{Bia89},
by time-of-flight and $B\rho$ measurements to determine their mass
over charge ratio $A/Q$, and the energy loss in an ionisation chamber
to determine their charge.
A typical resulting identification matrix is shown in
figure~\ref{fig:pid}.
To detect $\gamma$ rays, 74 BaF$_2$ detectors of the ``Ch\^ateau de
Cristal''~\cite{Bec84} and three small EXOGAM Ge clover
detectors~\cite{She99} were placed around the target.

To improve the particle identification in SPEG, not only the signal
from the ionisation chamber but also the charge deposited in the four
drift chambers (which are mainly used to determine particle positions)
can be used to determine the charge of the ions.
Similarly, for the $A/Q$ identification both the time-of-flight
measurement and the energy deposited in the beam-stopping scintillator
were used.
These four parameters were used as the $q_k$ for sorting and indexing:
the ionisation chamber signal as $q_1$, the time-of-flight signal as
$q_2$, the charge deposited in the drift chambers as $q_3$ and the
energy deposited in the scintillator as $q_4$, thus alternating
between mass and charge identification.
For each event, information from other detectors was stored along with
the $q_k$, e.g. time-of-flight information for the incoming secondary
beam, trigger information, the $\gamma$ ray energies and times of
detection measured by the BaF$_2$ detectors.

\begin{figure}
  \centering
  \includegraphics[width=6cm]{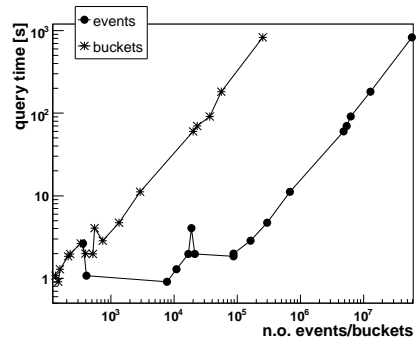}
  \caption{%%
    Query times selecting nuclei produced in the GANIL experiment
    as a function of the number of events (dots) and buckets (stars).
    The times were obtained on the author's analysis computer.
    They depend strongly on the distribution of events into buckets,
    but are roughly linear both in the number of selected events (with
    an offset) and in the number of buckets.
  }
  \label{fig:times}
\end{figure}

Using the sorting software, around 250\,000 buckets were filled for
$5.8\times 10^7$ events in two steps: in the first step, the list-mode
data were read and in the second step, the granularity of buckets was
increased.
An overview of query times on the author's computer is given in
fig.~\ref{fig:times}.
As can be seen, the query time is very small for selections of a small
number of events.
For $^{36}$Ca, the query time is slightly less than 2\,s.

\section{Summary}
\label{sec:summary}

A software has been developed which allows indexing and partial
sorting of data.
Query times for a subset of such an indexed dataset are roughly
proportional to the size of the queried subset.
This may be a large speed-up compared to reading time-ordered data.
The programs have proven to be very useful for the analysis of data
from an experiment on $^{36}$Ca performed at GANIL.

\appendix

\section{Acknowledgements}
\label{apx:ack}
The author wishes to thank W.~Korten, A.~G\"orgen (both CEA Saclay),
F.~Azaiez (IPN Orsay) and H.~H\"ubel (HISKP Bonn) for making possible
his stay at the CEA Saclay after the GANIL experiment and for
discussing this work.
Furthermore, he wants to thank all the contributors to the experiment
at GANIL for their support.

\end{document}